

\input harvmac

\def \ga {\alpha}
\def \gb {\beta}

\def \gp {\phi}
\def\gg {\gamma}
\def \gr {\rho}
\def \gd {\delta}
\def \gs {\sigma}

\def \bgb {\bar \beta }
\def \gij {\gg_{ij}}

\def \e#1 {{\rm e}^{#1}}
\def \const {{\rm const }}
\def \tf {{\tilde  f }}
\def \bbf {{\bar f}}

\def \ggij {{g_{ij}}}
\def \dg  {{\dot g}}
\def \ddg  {{ \ddot g}}
\def \df   {{\dot f}}
\def \ddf  {{\ddot f}}
\def \hg {{\hat g}}

\def\np {  Nucl. Phys. }
\def \pl { Phys. Lett. }
\def \mpl { Mod. Phys. Lett. }
\def \prl {Phys. Rev. Lett. }
\def \pr  { Phys. Rev. }

\Title{DAMTP - 92 - 26
\ \ \ hepth@xxx/9205058}
{\vbox{\centerline{A class of finite two - dimensional sigma models}
\vskip2pt\centerline{ and string vacua}}}

\centerline{ A.A. Tseytlin
\footnote{$^\dagger$}
{ e-mail: aat11@amtp.cam.ac.uk
\ \ \ On leave of absence from the Department of
Theoretical Physics, P. N. Lebedev Physics Institute, Moscow 117924, Russia.}}
\bigskip
\centerline{\it DAMTP}
\centerline{\it Cambridge University}
\centerline{\it  Cambridge CB3 9EW }
\centerline{\it United Kingdom }
\baselineskip=14pt plus 2pt minus 2pt
\vskip .3in

We consider a two - dimensional  Minkowski signature sigma  model with a
$2+N$ - dimensional target space metric  having a null Killing vector.
It is found that the model is finite to all orders of the loop expansion
if  the dependence of the ``transverse" part of the metric  $\ggij (u,x)$
on the light cone coordinate $u$  is subject to the standard renormalization
group  equation of the $N$ - dimensional sigma model,  $ {d\ggij\over du }=
\gb^g_{ij} =R_{ij} + ...\ $.
 In particular, we discuss the `one - coupling' case
when $\ggij(u,x)$ is a metric of  an  $N$ - dimensional symmetric space
$\gij(x)$
multiplied by a function  $f(u)$. The theory is finite if $f(u)$  is equal to
the
``running"  coupling of the symmetric space sigma model (with $u$ playing the
role of the RG ``time"). For example, the geometry of  space - time  with
$\gij$ being the metric of the $N$ - sphere is determined by the form of
the $\gb$ - function of the $O(N+1)$ model.  The ``asymptotic freedom" limit of
large $u$  corresponds to the weak coupling limit of small $2+N$ - dimensional
 curvature. We show that there exists a dilaton field which together with the
$2+N$ - dimensional metric solves the sigma model Weyl invariance conditions.
The resulting backgrounds thus represent  new tree level string vacua.
We remark on possible connections with some $2d$ quantum gravity models.

\Date{5/92} 

\baselineskip=20pt plus 2pt minus 2pt

1.  Only few examples of  exact solutions of the string
(tree--level) effective equations [1]   are explicitly
known at present. They include, in particular, group manifolds [2]  and plane
wave -- type spaces [3]. Below we shall present a new time - dependent solution
corresponding to a finite $2d$ sigma model. Like in the case of the spaces
considered in [3] the metric will have the Minkowski signature and  a null
covariantly constant Killing vector. However, the ``transverse"  part of
the metric will not be flat but will represent an arbitrary (for example,
symmetric) space. Also, the mechanism by which  conformal invariance conditions
will be satisfied will be novel.
 The $\gs$ - model divergences will not be absent automatically at each order
of the loop expansion but will vanish ``on - shell", i.e. one will be able to
 redefine  them away. The corresponding terms in the Weyl anomaly coefficients
($\bgb$ - functions ) will be cancelled by the contributions of the dilaton
and ``reparametrisation terms" [4].

Our model provides an example of how one can construct a conformal invariant
theory by  adding two extra  (one time - like and one space - like)
dimensions to a non - conformal one (in our case -- an arbitrary $N$ -
dimensional sigma model). This suggests a close connection with a particular
model of $2d$ scalar - tensor quantum gravity\foot{ For recent discussions of
scalar--tensor 2d quantum gravity from various points of view see e.g. [5--7].}
coupled to a  $\gs$ - model.\foot
{At the classical level similar
models (with symmetric target spaces) appear  as a result of dimensional
reduction  of higher dimensional Einstein theory  to $D=2$ (see e.g. [8]).} In
fact, a $2d$ gravitational system represented in the conformal gauge turns out
to be identical to the ``1-loop finite" form of our  $2+N$ - dimensional  $\gs$
-
model (with $N$ being the dimension of a symmetric space and
the conformal
factor and an extra scalar playing the role of the light cone  coordinates).
This correspondence implies that the properly modified `symmetric space $\gs$ -
model -- $2d$ gravity' model is also  finite to all orders in the loop
expansion.\foot  {Quantum properties of
supersymmetric extensions of such $2d$ gravity - matter models which can be
obtained e.g. by a dimensional reduction from $D=4$ supergravity (for an early
discussion of  r-trivial ``scalar - tensor" $D=2$ supergravity
models see [9])  were recently studied  in [10]. Our conclusion about
finiteness
of the corresponding  generalized  bosonic models  suggests that  similar
``modified" supersymmetric models may be finite  to all loop orders.
Let us note also that it should be  possible to construct a finite model  with
explicitly known metric  by  replacing the  ``transverse" part of the bosonic
action by an $N=2$  supersymmetric  sigma model with a homogeneous target
space.
The presence of the extra  two ``longitudinal" dimensions will make possible to
 get rid of the 1-loop divergences  preserving at the same time the higher loop
finiteness of the supersymmetric  ``sub - model".}

2. Let us  consider the following metric
$$ds^2 = G_{\mu \nu} dx^{\mu} dx^{\nu} =  -2dudv +  \ggij (u,x) dx^i dx^j   \ \
,
 \ \eqno {(1)}$$ $$ \ \ \mu , \nu = 0,1, ..., N, N+1 \ \ , \ \ \
  i,j = 1,...,N \ \ . $$
We shall  show that if the dependence of $\ggij $ on $u$ is subject to
a certain first order differential equation (nothing but the standard
RG equation of the $N$ - dimensional sigma model with the
metric $\ggij(u,x)$ and $u=\const$ playing the role of the RG ``time"
parameter) then  the $2+N$ - dimensional $\gs$ - model with the target space
metric (1) is UV finite.

As an important  example  we shall
consider  the `one - coupling' case when
 $$ \ggij(u,x) = f(u) \gij(x)    \ \ , \eqno (2) $$
where $\gij(x)$ will be assumed to be a metric of a symmetric (constant
curvature) space (one can of course consider a trivial generalization of (2)
 corresponding to a direct product of symmetric spaces with separate
``couplings" $f_n(u)$). Once the metric is chosen in the ``factorized" form the
restriction  that the $N$ - dimensional s,pace is symmetric  is necessary in
order to be able to represent the ``transverse" part of the  divergences in
terms of the original metric $\gij$ (i.e. to have renormalisability in terms of
one essential coupling constant $f$). The model will be finite under the
specific choice of $f$, i.e. for $f(u)$ equal to the ``running" coupling of the
symmetric space $\gs$ - model.

While all the basic formulas
will be true for a general $\ggij$ or a  symmetric
space $f\gij$ we shall  derive  the explicit perturbative  expressions
(eqs.(14),(24),(30)--(32), etc)  in the case  when $\gij$ corresponds to a
maximally symmetric $N$ - space with the  curvature tensor
 $$R_{ijkl}= {k \over (N-1) }  ( \gg_{ik} \gg_{jl} - \gg_{il} \gg_{jk} )
 \ \ , \ \  k\equiv {R\over N} = \const \ \ . \  \eqno (3) $$
It is useful first to give a
heuristic ``proof" of  the finiteness of the model (1),(2). The
path integral over $v$ produces a delta - function which  constrains $u$ to  be
constant. Then we are left with  the standard $\gs$ - model for $x^i$ (with
$f^{-1}$ playing the role of a constant coupling or a Planck constant ).
The corresponding divergences are proportional to    $$\int d^2 z \  T_{ij}
(x,f)\del_a x^{i} \del^a x^{j} \ \ , $$ where $T_{ij}$ is constructed from the
curvature tensor of $\gij$  and its derivatives [11]. If
$\ggij$ corresponds to  a symmetric space  $T_{ij}$ is proportional to $\gij$
with a constant coefficient. The final step is
to note that  the classical $u$ - equation of motion implies that $\gij \del_a
x^{i} \del^a x^{j}$ is proportional to the total derivative term $\del^2 v$
($u=\const$ according to $v$ - equation of motion). That is why  all
 the divergences are absent ``on  shell".

 This ``proof"  does not seem to depend on a form of $f(u)$.
\foot{It is not straightforward to generalize the above argument to the case
when  $\ggij$  is arbitrary, i.e. when the condition (2)  is relaxed. Given
that  the $2d$ sigma model is renormalisable in the generalized sense [12,11]
we
can still represent the counterterms in terms of the ``classical" metric  with
shifted $u$ (i.e. in terms of renormalized metric). However, the shifted $u$
depends on $x$ and hence is no longer constant. In contrast to the
factorized case (2), to achieve finiteness  even on the $u=\const$ background
we
do need to subject $\ggij(u,x)$ to a (differential) constraint.}
 What it actually proves,
however, is not the  finiteness of the model but the absence of divergences on
a
particular solution ($u=\const $) of the classical equations.  As it is easy to
understand, the latter property does not by itself guarantee that the
divergences are proportional to the classical equations of motion and hence can
be eliminated by  redefinitions of the $2d$ fields $u(z),\ v(z), \
x^i(z)$.\foot{This is related to the fact that $\del^2 u =0$  is a second order
equation, i.e. its general solution is given by $u(z) = u_0 + b_a z^a$. }

To establish the finiteness of the model on a flat $2d$ background one should
check that  the $\gb$ - function  for the  ``full" $\ \gs$ - model with the
target space metric $G_{\mu \nu }$ (1)  vanishes up to a ``reparametrisation"
term [12], i.e.
     $$ \gb^G_{\mu \nu} + 2 D_{( \mu} M_{\nu )} =0 \ . \eqno (4) $$
As we shall show,  (4)  is satisfied only for a particular  $\ggij(x,u)$ as
a function of $u$ (i.e.
for a particular $f(u)$ in the case of (2)). Moreover, the resulting metric (1)
 together with an
appropriate  dilaton background will represent a solution of the  Weyl
invariance conditions of the $\gs$ - model on a curved $2d$ background (and
thus
also a solution of the  string effective equations).

 The Weyl invariance conditions for the $\gs$ - model
$$ I= {1\over { 4 \pi  \ga'}} \int d^2 z \sqrt {g} [\  G_{\mu \nu
}(x) \del_a x^{\mu} \del^a x^{\nu}  +  \ga' R^{(2)} \gp (x)\ ] \ , $$
have the following general structure [1,4]
$$ {{\bar \gb}^G}_{\mu \nu } =
\gb^G_{\mu \nu} + 2\ga'  D_{\mu} D_{\nu} \phi + D_{( \mu} W_{\nu )} =0 \ ,
\ \ \eqno {(5)} $$
$$ \bgb^{\phi} = \gb^\phi + \ga' (\del_\mu \phi)^2 + \ha W^{\mu} \del_{\mu}\phi
=0  \ \ . \ \eqno {(6)} $$
The leading terms in the $\gb$ - functions and $W_\mu$ are well
known (in the dimensional regularisation and minimal subtraction
scheme) [11,12,1,4,13,14] $$ \gb^G_{\mu \nu}= \ga' R_{\mu \nu} + {1\over 2}
{\ga'}^2 R_{\mu \ga \gb \gg} R_{\nu}^{\ga \gb \gg}
+ {1\over 16} {\ga'}^3 [8 R_{\ga \gr \gs \gb} R_{\mu}^{\ \gs \gr \gg}
{R_{\nu}^{\ \ \ga\gb}}_{\gg} - 6 R_{\mu \ga \gb \nu } R^{\ga \gs \gr \kappa}
R^{\gb}_{\ \ \gs \gr \kappa} $$ $$ + 2 D_\gr R_{\mu \ga \gb \gg} D^{\gr}
R_{\nu}^{\ \ \ga \gb \gg}  - D_\mu R_{\ga \gb \gg \gr }
 D_\nu R^{\ga \gb \gg \gr}  ] + O(\ga'^4)  \ \ ,\eqno {(7)}$$
$$ \gb^\phi = {1\over 6 } (2+N - 26) -\ha \ga' D^2 \phi + {1\over 16}{\ga'}^2
R_{\mu \ga \gb \gg}R^{\mu \ga \gb \gg} + O(\ga'^3) \ \ , \eqno (8) $$
$$ \ \  W_\mu = {1\over 32} {\ga'}^3 \del_\mu ( R_{\ga \gb \gg \gr}
R^{ \ga \gb \gg \gr} )  + O(\ga'^4) \ \ . \eqno {(9)} $$
If (5) has a solution,  the finiteness condition (4) is satisfied for $M_{\mu}=
\ga'\del_\mu \phi + \ha W_{\mu}, \ $ i.e. a Weyl invariant $\gs$ - model is
also
UV finite on a  flat $2d$ background (the opposite may not necessarily be
true [15,4]).

The non-vanishing components of the connection and
 curvature of the metric $G_{\mu \nu }$  (1) are (we shall finally
specify all the formulas to the case of the metric (2))
 $$ {\hat \Gamma}^i_{jk} = { \Gamma}^i_{jk}\ \ ,
\ \ {\hat \Gamma}^v_{ij}=\ha \dg_{ij} = \ha \df\gij \ \ , \ \ {\hat
\Gamma}^i_{ju}=\ha g^{ik}\dg_{kj}  =
 \ha f^{-1}\df \delta^i_j \ \ ,\eqno (10)  $$ $$
 \dg_{ij}\equiv{d\ggij \over du}  \ \   ,\ \ \ \df\equiv{df\over du} \ \ , $$
 $$ {\hat R}^i_{jkl} = { R}^i_{jkl}\ \ , \ \
{\hat R}^v_{iju}\equiv H_{ij}
\ \ , \ \ {\hat R}^i_{uju}= g^{ik} H_{kj}\ \ , \ \
$$ $$ H_{ij}=-{1\over 2}  (\ddg_{ij}-\ha  g^{mn}\dg_{im}\dg_{nj}) =
-{1\over 2} (\ddf-\ha  f^{-1}\df^2) \gij \ \ , \eqno (11) $$
$${\hat R}^v_{ijk} = -{\hat R}_{uijk}\equiv F_{ijk}\ \ , \ \
 F_{ijk}= \ha [\del_j \dg_{ik} + \dg_{jm} { \Gamma}^m_{ik} - (j \leftrightarrow
 k) ] =\ha ( g_{km} { {\dot \Gamma}}^m_{ij} -
g_{jm} { {\dot \Gamma}}^m_{ik}) \ \ , \eqno (12) $$
$$F_{ijk} (\ggij = f\gij) =0 \ \ .  $$
For a general $\ggij$ the   non-zero elements of  $\gb^G_{\mu \nu} $
(7)  are $\gb^G_{ij} $ , $\gb^G_{uu}$ and $\gb^G_{iu}$ (the latter vanishes in
the factorized case (2)). In general they are non-trivial functions of $u$ and
$x^i$. It is easy to understand that there are no non-vanishing contractions
of the curvature components with $u,v$ indices.  $\gb^G_{ij}$  depends only
on ${ R}^i_{jkl} , \ $  its covariant derivatives  and $ \ggij$ and therefore
coincides with the $\gb$ - function of the  $\gs$ - model with the metric
$\ggij$.   In the special case of (2),(3)
 $f^{-1}$ plays the role of the coupling of the symmetric space $\gs$ - model
(we
may absorb the scale of $\gij$  into $f$  making $k$ in (3) equal, for example,
 +1 or --1 depending on a sign of the curvature $R$).  As a result, we find for
$\gb^G_{ij}$ (see (7))
   $$\gb^G_{ij} = \gb^g_{ij} = \gb (f) \gij \ \ , \eqno (13) $$
 $$ \gb^g_{ij} = \ga' R_{ij} + O(\ga'^2) \ \ , \ \ \ \gb = f
\sum_{n=1}^{\infty}
c_n (af^{-1})^n \ \  , $$
 $$ \gb(f) = a +  (N-1)^{-1}a^2f^{-1}  + {1 \over 4} (N-1)^{-2}(N+3)a^3f^{-2} +
 O(a^4 f^{-3})   \ \  , \eqno (14) $$
$$a \equiv  \ga' k \ \ .  $$
 The leading terms  in  $\gb^G_{uu}  $ and $\gb^G_{iu}$ are given by (see
(7),(11),(12))\foot{ In the symmetric space case (2) $\gb^G_{uu}$   does not
receive the 2 - loop (and certain higher loop) contributions if the $\gb^G$ -
function is computed in the `dimensional regularisation plus minimal
subtraction' scheme (i.e. if there are no  terms in $\gb^G_{\mu \nu}$ with
Ricci
tensors as factors). However,  there $are$  3 - and higher loop contributions
to
it. For example,  there is the   3 - loop  term in  (13)
 $ -{1\over 4}\ga'^3 f^{-4} \df^2 R_{ijkl}R^{ijkl}  $  which originates from
the
derivative   $D_\mu R_{\ga \gb \gg \gr }D_\nu R^{\ga \gb \gg \gr}$ -  term in
(7).}
$$\gb^G_{uu} = \ga'R_{uu} +  \ha \ga'^2 R_{uijk} R_{u}^{ijk} + O(\ga'^3)=
\ga' g^{ij} H_{ij}  +  \ha \ga'^2 F_{ijk} F^{ijk} + O(\ga'^3) \ \ ,
$$ $$ \gb^G_{uu} =  -{1\over 2}\ga' Nf^{-1} (\ddf-\ha
f^{-1}\df^2)  + O(\ga'^3) \ \ , \ \eqno (15) $$
$$ \gb^G_{iu} = \ga' g^{jk}F_{jki}  -  \ha \ga'^2 F_{mnk} R_{i}^{mnk} +
O(\ga'^3)\ \ , \ \  \gb^G_{iu} ( \ggij= f\gij) =0  \ \ . $$
Let us now try to find such vectors $M_\mu$ which may satisfy the finiteness
condition (4). Using (13) and  that $\gb^G_{iv}=0,\ \gb^G_{uv}=0 \    $ we can
get the following equations (the covariant derivatives are now defined
with respect to $\ggij$, see (10))
$$ \gb^g_{ij} +  2D_{( i} M_{j )} - 2{\hat \Gamma}^v_{ij}M_v =0 \ \ ,
\ \ \  $$
i.e.
$$  \bgb^g_{ij} = \dg_{ij} M_v  \ \ ,
 \ \ \ \ \bgb^g_{ij}\equiv \gb^g_{ij} +  2D_{( i} M_{j )} \ \ ,
 \eqno (16) $$
$$ \gb^G_{uu}  = -2 \del_u M_u  \ \ ,  \eqno (17) $$
$$\gb^G_{iu} = - \del_i M_u - \del_u M_i + g^{jk}\dg_{ij}M_k \ \ , \eqno (18)
$$ $$ \del_i M_v + \del_v M_i =0  \ \ , \ \ \
\del_u M_v + \del_v M_u =0  \ \ .  \eqno (19) $$
 Since all the components of $\gb^G_{\mu \nu}$ do not depend on $v$,
the only $v$ - dependence that is possible in $M_\mu$ is a linear term in
$M_u$. Then the general solution of (19) is given by
  $$ M_v=mu + p  \ \ ,  \  \ \  M_u = -mv + Q(u,x) \ \ , \ \ \
M_i = M_i(u,x) \ \ , \ \ \ \ p\ ,\  m =\const \ \  . \eqno (20) $$
For a given $\ggij(u,x) \ $  $ \gb^G_{uu} $ and $  \gb^G_{iu} $ are some
given  $N+1$ functions of $u$ and $ x$  so that   one can always  satisfy  the
 equations (17) and (18)  by properly chosen $N+1$ functions  $M_u$  and
$M_i$ (once we solved (17), we can put (18) in the form $ \del_u M_i +
h_i^j(u,x) M_j = E_i(u,x) $ which always has a solution).

Having determined $M_u$ and $M_i$ as functionals of $\ggij$ we  are left with
the
final equation (16). It should be interpreted as an equation for $\ggij(u,x)$.
Using (20) and  introducing\foot{We are  considering non-zero  $m$ for
generality; to get a Weyl invariant model one must actually set $m=0$ (see
below).} $$\tau = m^{-1} \ln  (mu+p) \ \ ,  \ \ m\not=0 \ \ ; $$ $$
 \ \ \ \ \tau = p^{-1}u  \ \ , \ \ m=0 \ \ ,$$
 we can  represent (16) in the form
$$ {d\ggij \over d \tau } = \bgb^g_{ij}  \ \ . \  \eqno (21) $$
This concludes the proof of the statement  that if the  metric $\ggij$
depends on $u$ in such a way that it satisfies the standard RG equation of  the
$N$
- dimensional sigma model (with some particular
reparametrisation vectors) then the $2+N$ - dimensional model based on (1) is
UV
finite to  all orders of the loop expansion.

The above  argument for finiteness   is  simplified  in  the particular case of
symmetric space (2). Since $\gb^G_{iu}=0$ and scalar functions (e.g.
$ \gb^G_{uu}$) are $x$ - independent
we set $M_i=0, \ \del_iM_u=0$ and thus  solve (16),(17),(18) by (see
(13),(14))
$$ M_v=mu + p  \ \ ,  \  \ \  M_u = -mv + Q(u) \ \ ,$$
$$  {\dot Q} ={1\over 4}\ga' N(f^{-1}\ddf- ha  f^{-2}\df^2)  + O(\ga'^3) \ \ ,
 \eqno (22) $$
 $$ M_v\df \gij = \gb (f) \gij\ \ ,  \ \ \ $$
i.e.
$${df\over d\tau } =\gb(f) \ \ . \eqno (23) $$
 Eq.(23) with $\gb$ given by (14) has the obvious perturbative solution
 $$
f(u) = a[ \tau + (N-1)^{-1} \ln  \tau + O(\tau^{-1})]  \ \  . \ \eqno (24) $$
Having found $f(u)$ from (19)  one determines $Q$  from (18) by integration.
This proves the existence of a finite  sigma model based on
(1),(2).

3. To solve the Weyl invariance conditions (5),(6)  one needs to  establish
that
$M_\mu$ in (4)  can be represented in the form
$$M_{\mu}=\ga'\del_\mu \phi + \ha W_{\mu}\ \ . \eqno (25) $$
Since it is known [1,4] that $W_\mu $ vanishes in the one -  and  two - loop
 approximation\foot{In general, $W_\mu$ is constructed from the curvature
and its covariant derivatives and thus  depends only on $u$.}  $m$ in
(20) must vanish in order for a solution to exist. In fact, eq.(19) is
consistent with $M_\mu$ being a gradient only if $ \del_vM_u=0$. We find
from (17),(18) ( in the 2 - loop approximation)
$$ \gb^G_{uu}  = -2\ga'\del^2_u \phi + O(\ga'^3) \ \ ,   $$
$$\gb^G_{iu} = - 2\ga'\del_i \del_u \phi + g^{jk}\dg_{ij}\del_k \phi +
 O(\ga'^3) \ \ . $$
Using the  expression for $ \gb^G_{\mu \nu}$ it is possible to check that the
integrability condition for these equations is satisfied, i.e. there exists a
solution
$$  \phi = \ga'^{-1} [  pv   +  F(u,x)] \ \ , \ \ \ \ p=\const \ \ ,  \eqno
(26)
$$ which together with the metric (1) satisfies (5),(6). At the higher loop
level  the expression for $M_\mu$ found in the previous section  will be
``distributed" between the $\phi$ - and $W_\mu$ - terms in (25).

All is more transparent in the symmetric space case (2) where  the
differential equations become the  ordinary ones and their  integrability is
obvious.  We get
 $$ \phi = \ga'^{-1} [ pv + F(u)  ] \ \ \ , $$  $$ {\dot F}= Q
-\ha W_u   \ \ , \ \ \ \  W_v =0  \ \ , \eqno (27) $$   To  find $Q$, $W_u$
and $\phi$ from  (22), (9)  and (27) it is useful to  change the integration
variable from $u$ to $f$ using the ``RG equation" (23)
$$  Q= q + {1\over 4}\ga'
N p^{-1} \int^f_\infty d\bbf [ \bbf^{-1} { d\gb \over d\bbf}  -\ha \bbf^{-2}
\gb
(\bbf) ]  + O(\ga'^3)  \ \ , \ \  \ \ q= \const \ \ , \eqno (28) $$ $$ W_u =
-{1\over 8}\ga' a^2 N(N-1)^{-1} f^{-3}\df + O(\ga'^4) =  -{1\over 8}\ga' a^3
p^{-1}N(N-1)^{-1} f^{-3} + O(f^{-4})  \ \ , \ $$
$$  \phi = \ga'^{-1} ( qu + pv ) \  + \
 \int^f_\infty d\bbf \gb^{-1}(\bbf)\  \{ \ {1\over 4} N \int^\bbf_\infty  d\tf
\ [ \tf^{-1} { d\gb \over d\tf} $$ $$  -\ha \tf^{-2} \gb (\tf)  +  O(\ga'^3) ]
-
\ha \ga' p W_u \} \ \ .
 \eqno (29) $$
Using the expression for $\gb(f)$ (14) we get  the following
large $u$ expansions in powers of the ``coupling" $f^{-1}$
$$ f(u) = a[p^{-1} u + (N-1)^{-1} \ln
(p^{-1} u ) + O(u^{-1}\ln u )]  \ \  , \eqno (30) $$
$$  Q= q + {1\over 16}\ga' N
p^{-1}[ 2af^{-1} + 3 a^2 (N-1)^{-1} f^{-2} + O(f^{-3})  ] $$  $$ = q +
\ga'[{1\over 8} N u^{-1}  + O(u^{-2}\ln u )]\ \ , \ \  \  \
  W_u= O(u^{-3})\ \ , \eqno (31)
$$ $$ \phi =\phi_0 + \ga'^{-1} (qu + pv) + {1\over 8} N\ [ \ln f -
a(N-1)^{-1}f^{-1} +O(f^{-2}) ] $$
$$ = \phi_0 + \ga'^{-1} (qu + pv) + {1\over 8} N
\ln (ap^{-1}u) + O(u^{-1}\ln u ) \ \ . \eqno (32) $$

The resulting dilaton contribution
to $ \bgb^{\phi}$  (eqs.(6),(8)) is given by
$$ \Delta \bgb^{\phi}=-\ha \ga'
D^2 \phi  +\ga' (\del_\mu \phi)^2 + \ha W^{\mu} \del_{\mu}\phi = {1\over 4}
pNf^{-1}\df - 2 \ga'^{-1} pQ + {1\over 2} pW_u  \ \ . \eqno (33) $$
It cancels  the contribution of the higher loop $\phi$ - independent terms in
$\gb^\phi$ (8) so that the central charge of the model $\bgb^{\phi}$ is  equal
to that of the free  $N+2$ - dimensional theory plus the contribution of the
linear terms in the dilaton. To prove this one is to note that  once (5)  is
satisfied $\bgb^\phi$ is constant [16]
 and hence  can be computed at any value of
$u$, e.g. $u=\infty$. Given that all higher loop contributions should vanish in
the weak coupling limit of large $u$ it is sufficient to  compute $\bgb^{\phi}$
in the leading order  approximation. Substituting (31),(32) into (33) we get
$$\Delta \bgb^{\phi}= -2\ga'^{-1}pq -{1\over 8} N
(N-1)^{-1}a^2 f^{-2} + O(f^{-3}) \ \ . \eqno (34) $$
The 1 - loop $O(f^{-1})$ terms in (34) cancelled automatically while the 2 -
loop  $O(f^{-2})$ term  cancels against the contribution of the $R^2$ term
in (8), etc. The final expression for the central charge coefficient $
\bgb^\phi
$  is    $$ \bgb^\phi = {1\over 6 } (N - 24) - 2\ga'^{-1}pq  \ \ . \eqno (35)
$$
Thus one can satisfy the zero central charge condition (6) for arbitrary $N$
by a proper choice of the constants $p$ and $q$.\foot{We are assuming that
 $u, \ v, \
x^i $  have dimension +1 and $G_{\mu \nu}, \ \gij, \ \phi, \ f , \ \gb$ are
dimensionless. Then  $ [\ga']=2, \ [p]=[q]=1, \ [a]=0, \ [k] =-2, \ [Q]=1,\  $
etc.  Since $p$ is a free
parameter we may set it, for example, equal to $\sqrt {\ga'}$ (or 1 if
$\ga'=1$). Then $\tau$ in (24) becomes  equal simply to $u$. }

4. We  have found the standard (first order)
RG  equation of an $N$ - dimensional theory  (21),(23)  in the process of
solving
the (second order) finiteness conditions (4) of a higher $N+2$ -
dimensional model. This  is  reminiscent of  the   ``semiclassical"
correspondence between the conformal invariance conditions in an  $N+1$ -
dimensional theory  and the RG equations in  an  $N$ - dimensional one. The
latter  relation can be  established in the   presence of an (asymptotically)
linear dilaton background (see e.g. [17] and Appendix C of ref.[18]). In fact,
consider a generic equation for a  ``massless" (marginal perturbation) coupling
$\psi$   ( $f$ in (1),(2) is an example of such
coupling)
 $$ \bgb^\psi = \gb  -\ha \ga' D^2 \psi +  \ga' D^{\mu}
\phi\del_\mu \psi =0 \ \  , \ \  \ \
\gb = \gb (\psi) ={\del V \over \del \psi} \  \ . \eqno (36) $$
If the $N+1$ - dimensional metric  and the dilaton  are given by
$ds^2 = - b dt^2 + ds^2_N \ , \ \ \phi = \phi_0 + bt \ \  ( b = \const )$  and
$\psi$ depends only on $t$  then in the limit of $b \rightarrow \infty$ eq.(36)
reduces to  ${d\psi \over dt} = \gb $   (we  set  $\ga'=1$). If, on the
other hand,  we  use the $N+2$ - dimensional metric (1)   and the dilaton
(26) which is linear in $v$ and assume that $\psi=\psi (u)$  then (36)  reduces
to (23), i.e.  $ \ \ p {d\psi \over du} = \gb $.

The ``1-loop finite" form of the symmetric case metric (1),(2),(30) is
particularly simple (we set $a=p=1$)
 $$ds^2  =  -2dudv +  u \gij (x) dx^i dx^j   \ \  .\ \eqno {(37)}$$
It is closely related to the following $2d$ gravity model
$$ S = - \int d^2 z {\sqrt \hg } \  \e{-2\phi}   \ [  \ {\hat R } \ + 4 (\del
\phi )^2  -    \gij (x)\del_a x^{i} \del^a x^{j}  \ ] \ . \eqno (38)  $$
In the conformal gauge $ \hg_{ab}= {\rm e }^{2\rho}\delta_{ab}$ one can put
(33)
into the  $2+N$ - dimensional sigma model form by identifying
$$ u=\e{-2\phi} \ \ , \ \  \ \ v = \rho + \ha \e{-2\phi} \ \ . \eqno (39) $$
To get a  model which is finite to all loop orders one should modify (38) by
replacing the coefficient $u$ of the $\gs$ - model part of (38) by
the full $f(u)$.

The fact that $u$ plays the role of an RG ``time" parameter  (equal to
logarithm of a $2d$ cutoff)  which in the  case of a covariant regularisation
is coupled   to the conformal factor of  a $2d$ metric suggest  a connection
with a different `$2d$ quantum gravity  -- sigma model' theory
$$ S =  \int d^2 z {\sqrt \hg }    \ [ - \ v {\hat R }\
+    \ggij (\tau , x)\del_a x^{i} \del^a x^{j}  \ ] \ . \eqno (40)  $$
Here $v(z)$ is a scalar field  and  the $2d$ metric in the conformal gauge is
$$\hg_{ab}= {\rm e }^{2u}\gd_{ab} \ \ , \ \ \ u=u(z)   \ \ \eqno (41) $$
(note that these identifications are opposite to what was assumed in
(38),(39)).
The argument $\tau$ of  $\ggij$ in (40)  indicates the dependence of the
coupling on  an RG parameter  on a flat $2d$ background. It is effectively
replaced by $u(z)$ once  a covariant cutoff is used (see e.g. [19]).
Then the statement that the model (36) is finite if $\ggij$ depends on $u$
according to the RG equation is  similar to what one expects to find in   the
$2d$ quantum gravity context (cf.[20,17,19]).\foot{Though an interpretation of
$v$ remains obscure let us mention that the presence of both $u$ and $v$ may be
related to the  presence of two string world sheet coordinates.}

 In conclusion, let us  mention some particular cases. If the metric $\gij$ is
flat
(this includes, for example, the general case of $N=1$, i.e. $D\equiv 2+N =3$)
then $\gb$ in (14) is zero (but $\gb^G_{uu}$ in (15) is non-trivial). Eq.(23)
implies that we should set $p=0$ ($\df=0$ gives a flat space solution).
As a consequence, we get a finite and Weyl invariant model  for an
$arbitrary$  $f(u)$ (with $Q$ and the dilaton $\phi$ being defined by (22) and
(26) and $W_\mu=0$). Note that this $2+N$ - dimensional space has a
conformally flat metric (1) and  a non - vanishing curvature  (11).

The model with $D=4$ (i.e. $N=2$) is represented by (1),(2) with $f(u)$ being
the running coupling of the $O(3)$ sigma model  (we assume that $k$ in (3) is
positive,
i.e. the metric is  conformal
to that of $R^2 \times S^2$). As a result,
the geometry of this $D=4$ space - time is determined by the behaviour
of the $\gb$ - function of the $O(3)$ $\ \gs$ - model.

\bigskip

\bigskip
I am grateful to E. Martinec for  suggesting  that the  statement of
the original version of this paper (which was restricted to  the symmetric
space sigma models)  should be true also in  the `multi - coupling' case. This
prompted the revision of the paper. I would like also  to acknowledge J. Russo
for discussions of $2d$  gravity models.  Part
of this work was done during the visit to the University of Adelaide and  I
would
like to thank the members of the theory group there, in particular J. McCarthy,
S. Poletti and D. Wiltshire for a hospitality and useful discussions.
 I am grateful to Trinity College, Cambridge for a financial support through a
visiting fellowship.

\vfill\eject

\centerline{\bf References}
\bigskip

\item {[1]} C. Lovelace, \pl B165(1986)409 ; \np B273(1986)413 ;

   E.S. Fradkin and A.A. Tseytlin, Phys. Lett. B158(1985)316 ;

   Nucl. Phys. B261(1985)1 ;

 C.G. Callan, D. Friedan, E. Martinec and M.J. Perry, \np B262(1985)593 .

\item {[2]} E. Witten, Commun. Math. Phys. 92(1984)455 ;

E. Braaten, T.L. Curtright and C.K. Zachos, \np B260(1985)630 ;

S. Mukhi, \pl B162(1985)345 .

\item {[3]} D. Amati and C. Klimcik, \pl B219(1989)443 ;

G. Horowitz and A.R. Steif, Phys. Rev. Lett. 64(1990)260 ;

Phys. Rev. D42(1990)1950;

R.E. Rudd, \np B352(1991)489 .

\item {[4]}  A.A. Tseytlin, \pl B178(1986)34 ; \np B294(1987)383 ;

Int. J. Mod. Phys. A4(1989)1257 .;

G.M. Shore, \np B286(1987)349 ;

H. Osborn, \np B294(1987)595 .

\item{[5]} R. Jackiw, in: Quantum Theory of Gravity, ed. S.Christensen (Adam

Hilger, Bristol 1984) ;

 A.H. Chamseddine, Phys. Lett. B256(1991)2930; \np B368(1992)98 ;

T.T. Burwick and A.H. Chamseddine, preprint ZU-TH-4/92 ;

T. Banks and M. O'Loughlin, Nucl. Phys. B362(1991)649.

\item{[6]}C.G. Callan, S.B. Giddings, J.A. Harvey and
A. Strominger,

\pr D45(1992)1005.

\item{[7]} J. Russo and A.A. Tseytlin, preprint DAMTP-1-1992 / SU-ITP-92-2 ;

H. Verlinde, preprint PUPT-1303 ;

A. Strominger, preprint UCSBTH-92-18 .

\item{[8]} P. Breitenlohner, D. Maison and G. Gibbons, Commun. Math. Phys.
120(1988)295.

\item{[9]} E.S. Fradkin  and A.A. Tseytlin, \pl B106(1981)63 .

\item{[10]} B. de Wit, M.T. Grisaru, E. Rabinovici and H. Nicolai,

preprint CERN-TH-6477/92 .

\item{[11]} J. Honerkamp, \np B36(1972)130 .

\item{[12]} D. Friedan, \prl 51(1980)334 ; Ann. Phys. 163(1985)318 .

\item{[13]} S.J. Graham, Phys. Lett. B197(1987)543 ;

     A.P. Foakes and N. Mohammedi, Phys. Lett. B198(1987)359 .

\item{[14]} I. Jack, D.R.T. Jones and N. Mohammedi, Nucl. Phys.
   B322(1989)431;

  Nucl.
Phys. B332(1990)333

\item{[15]} C.M. Hull and P.K. Townsend, \np B274(1986)349 .

\item{[16]} G. Curci and G. Paffuti, \np B268(1987)399 .

\item{[17]} S.Das, A.Dhar and S. Wadia, \mpl A5(1990)799 ;

 A. Cooper, L. Susskind and L. Thorlacius, \np B363(1991)132 ;

 A. Polyakov, Princeton  preprint PUPT-1289 (1991) .

\item{[18]} A.A. Tseytlin, preprint DAMTP-15-1992 .

\item{[19]}  A.A. Tseytlin, Int. J. Mod. Phys. A5(1990)1833 .

\item{[20]} F. David, Mod. Phys. Lett. A3(1988)1651 ;

     J. Distler and H. Kawai, Nucl.Phys. B321(1989)509 ;

     S.R. Das, S. Nair and S.R. Wadia, Mod.Phys.Lett. A4(1989)1033 ;

     J. Polchinski, Nucl. Phys. B324 (1989)123;

  E.J. Martinec, Lectures at the 1991 Trieste Spring School on

String Theory and Quantum Geometry, Trieste, Italy, Apr 15-23, 1991,

 preprint RU-91-51 .

 \vfill
\eject
\bye